\documentclass[12pt,preprint]{aastex}

\newcommand\litl{\rm\scriptscriptstyle}
\newcommand{\expo}[1]{\ensuremath{10^{#1}}}
\newcommand{\eg}{{\rm e.g.},}

\newcommand{\ie}{{\rm i.e.},}
\renewcommand{\micron}{\ensuremath{\,\mu{\rm m}}}

\newcommand\pc{\ensuremath{\,{\rm  pc}}}
\newcommand\av{\ensuremath{\,{\rm  A_v}}}
\newcommand\mjysr{\ensuremath{\,{\rm MJy}\,{\rm sr}^{-1}}}

\newcommand\pdeg{\ensuremath{.\negthinspace^{\circ}}}
\newcommand\hi{{\sc H~i}}
\newcommand\spitzer{{\sl Spitzer}}
\newcommand\iras{{\sl IRAS}}

\slugcomment{Accepted for publication in vol 154 of Astrophys. J. Supp. Ser.}

\shorttitle{Structure in Galactic First Look Survey}
\shortauthors{Ingalls et al.}

\begin{document}


\title{Structure and Colors of Diffuse Emission in the Spitzer Galactic
First Look Survey}

\author{James G. Ingalls\altaffilmark{1},
  M. -A. Miville-Desch\^enes\altaffilmark{2}, William
  T. Reach\altaffilmark{1}, A. Noriega-Crespo\altaffilmark{1}, Sean
  J. Carey\altaffilmark{1}, F. Boulanger\altaffilmark{3},
  S. R. Stolovy\altaffilmark{1}, Deborah L. Padgett\altaffilmark{1}, 
  M. J. Burgdorf\altaffilmark{1},
  S. B. Fajardo-Acosta\altaffilmark{1},
  W. J. Glaccum\altaffilmark{1}, G. Helou\altaffilmark{1},
  D. W. Hoard\altaffilmark{1}, J. Karr\altaffilmark{1}, J.
  O'Linger\altaffilmark{1}, L. M. Rebull\altaffilmark{1}, J.
  Rho\altaffilmark{1}, J. R. Stauffer\altaffilmark{1}, \&
  S. Wachter\altaffilmark{1}}

\altaffiltext{1}{Spitzer Space Telescope Science Center, California
  Institute of Technology, 1200 East California Boulevard, MS 220-6,
  Pasadena, CA 91125; Send offprint requests to J. Ingalls: ingalls@ipac.caltech.edu.}
\altaffiltext{2}{Canadian Institute for Theoretical Astrophysics, 60
  St-George Street, Toronto, Ontario, M5S 3H8, Canada.}
\altaffiltext{3}{Institut d'Astrophysique Spatiale, Universit\'e Paris-Sud, B\^at. 121, 91405, Orsay, France.}


\begin{abstract}
We investigate the density structure of the interstellar medium using
new high-resolution maps of the 8\micron, 24\micron, and 70\micron\
surface brightness towards a molecular cloud in the Gum Nebula, made
as part of the {\sl Spitzer Space Telescope} Galactic First Look
Survey.  The maps are correlated with 100\micron\ images measured
with \iras.  At 24 and 70\micron,
the spatial power spectrum of surface brightness follows a power law
with spectral index --3.5.  At 24\micron, the power law behavior is remarkably
consistent from the $\sim 0\pdeg 2$ size of our maps down
to the $\sim 5\arcsec$ spatial resolution.  Thus, the structure of the
24\micron\ emission is self-similar even at milliparsec scales. The
combined power spectrum produced from \spitzer\ 24\micron\ and \iras\ 25\micron\
images is consistent with a change in the power law exponent from
--2.6 at spatial wavenumber $k\sim 2\times\expo{-3} 
\,({\rm arcsec})^{-1}$ to --3.5 at $k\sim 4\times\expo{-3} 
\,({\rm arcsec})^{-1}$. The decrease may be due to the transition from
a two-dimensional to three-dimensional structure.  Under this
hypothesis, we estimate the thickness of the emitting medium to be
0.3\pc.


\end{abstract}





\section{Introduction}
More than fifty years ago, astronomers began to
reconsider the concept of a homogeneous density and velocity structure
for the interstellar medium (ISM).  The seminal lecture on astrophysical
turbulence by \citet{cha49} marked the recognition that studies of the
ISM were incomplete without knowledge of the physical and
observational consequences of turbulence. The structure and
the dynamics of the ISM affect critically its chemistry
\citep*{spa96,jou98,rol02} and its star formation capacity \citep[see
  for example the review by][]{mac04}.  The ramifications of a
turbulent velocity and density field for radiative transfer must be
taken into account when interpreting spectral line and continuum
observations of interstellar clouds
\citep{pad98,heg00,heg03,oss02,juv03}. Finally, knowledge of the
structure of Milky Way interstellar matter is essential to the proper
interpretation of extragalactic counts and measurements of the spatial
distribution of the Cosmic Microwave Background \citep{gau92}.

The structural statistics of the ISM are self-similar on a wide range of scales,
from hundreds of parsecs down to $\sim 0.02\pc$ \citep*{ben01}.  The angular power
spectrum of two-dimensional (2-D) images of interstellar emission and
absorption derived using a variety of tracers yields a power law as a function of 
wavenumber, $k^{\beta}$, with  exponent ranging from $\beta\sim -3.6$
\citep{mi03a} to $\beta\sim -2.5$ \citep{ben01} \citep*[see also the
  review by][]{fal04}.  
Power spectral analysis of the far-infrared
emission from the Galactic cirrus results in the same range of values
on angular scales of 1\arcmin\ and larger \citep{gau92,her98}.  

The {\sl Spitzer Space Telescope} allows us to extend the examination
of turbulent density fields down to $\sim 5\arcsec$ scales.  We
present new high-resolution maps of the 8\micron, 24\micron, and
70\micron\ diffuse emission towards the Gum Nebula,
made as part of the \spitzer\ Galactic First Look Survey (\S2).  In
\S3 we derive colors between the \spitzer\ and \iras\ bands and
perform a power spectral analysis of the \spitzer\ images.  For the
24\micron\ and 70\micron\ maps we find power law exponents similar to
those derived from \hi\ observations, with no detectable break at high
wavenumber down to spatial scales of $\sim 0.01\pc$. 
We discuss implications of our results for interstellar structure in \S4.  

\section{Observations}
\subsection{The {\sl Spitzer} Galactic First Look Survey}
The new data that we analyze form part of the {\sl Spitzer Space
Telescope} \citep{wer04} Galactic First Look Survey (GFLS)
\citep{nor04}\footnote{http://ssc.spitzer.caltech.edu/fls/galac/}.
We describe here observations made using the IRAC \citep{faz04} and
MIPS \citep{rie04} instruments on board \spitzer.  The IRAC
observations (PID 104; ads/sa.spitzer\#6579712)
were conducted on 2003 December 7, and the MIPS observations (PID 104;
ads/sa.spitzer\#6578176) were conducted on 2003 December 9.

\begin{figure}
\epsscale{.70}
\plotone{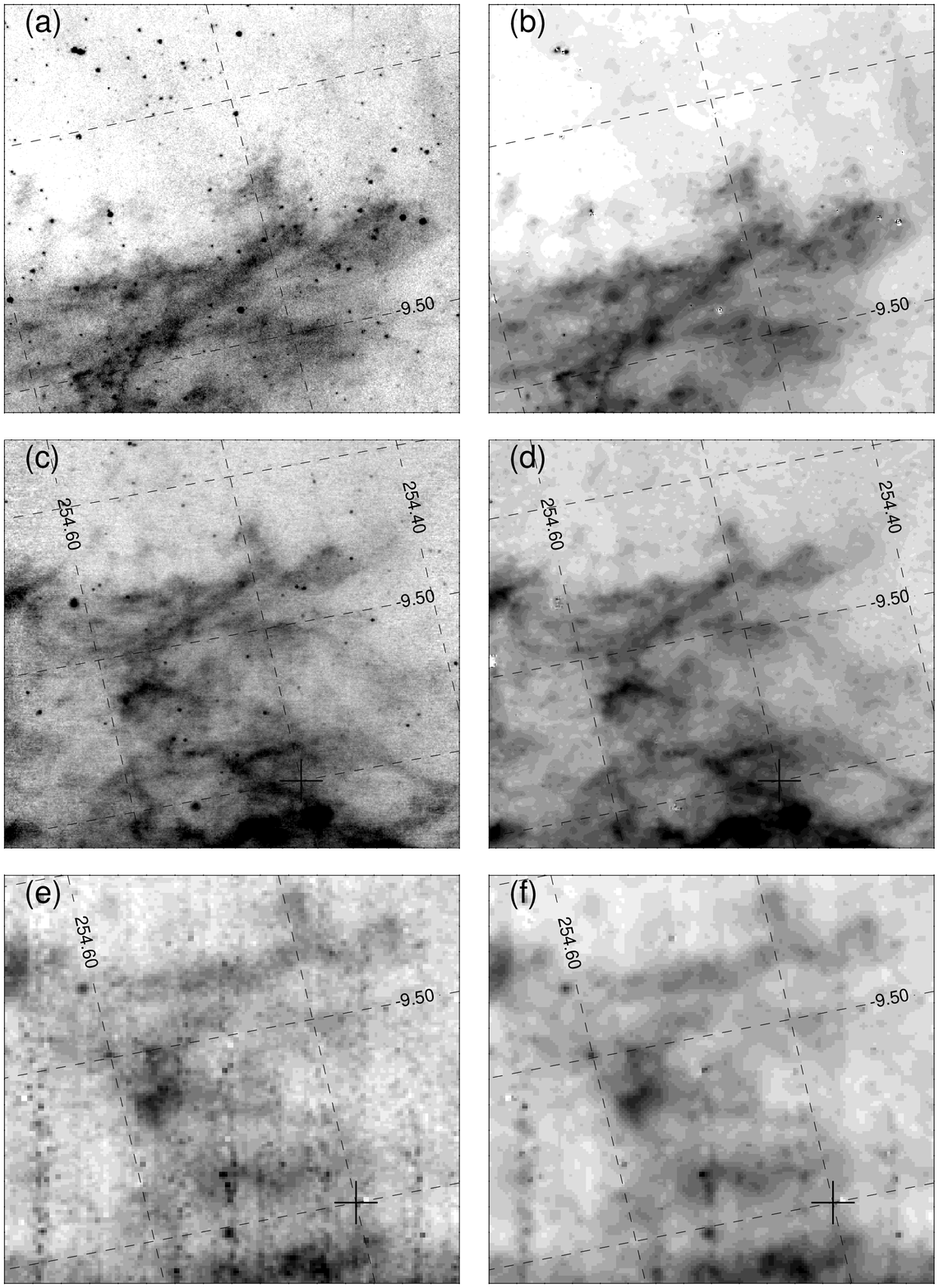}
\caption{ {\sl Spitzer} surface brightness maps used for power spectral
analysis. The field includes a molecular cloud associated with the Gum
Nebula.  ({\it (a)} and {\it (b)}): IRAC $8\micron$ gray levels range from 0.28
(white) to 0.33\mjysr (black).  ({\it (c)} and {\it (d)}): MIPS 24$\micron$ gray
levels range from 19.1 to 20.8 \mjysr.  ({\it (e)} and {\it (f)}): MIPS 70$\micron$
gray levels range from 20 to 490 \mjysr.  The left column---frames
{\it (a)}, {\it (c)}, and {\it (e)}---depicts the original post-BCD
mosaic images.  The right column---frames {\it (b)}, {\it (d)}, and
{\it (f)}---shows the results of point
source subtraction and noise filtering (see text) applied to the left
column images.  Each map has the same grid of Galactic coordinates
superimposed (grid lines are separated by 0.1 degree).  A cross in
each of {\it (c)} through {\it (f)} shows the position of cloud DC
254.5-9.6 \citep{har86}.  \label{fig1}}\notetoeditor{This figure
  should appear as a plate in the Journal}
\end{figure}

The surface brightness maps on which we have performed power spectral analysis 
are shown in
Figure \ref{fig1}.  The images on the left side of the figure are the output 
of the automated post-Basic
Calibrated Data (post-BCD) \spitzer\ calibration
pipeline\footnote{http://ssc.spitzer.caltech.edu/postbcd}.  For each
of the three \spitzer\ bands considered in this study---IRAC 8\micron\
and MIPS 24 and 70\micron
---we have analyzed slightly different fields, albeit with
considerable overlap.  There was a small mismatch
in the observed IRAC and MIPS fields caused by a $\sim 2\arcdeg$
rotation of \spitzer\ between the two observation dates.  We also
truncated the 70\micron\ mosaic due to excessive noise in portions of
the image.  

Each of the examined regions are centered on Galactic coordinates
$(l,b)=(254.5,-9.5)$ and cover a square of size $\sim 0\pdeg 1 -
0\pdeg 3$ (see Figure \ref{fig1}), comprising less than 2\% of the GFLS total
sky coverage.  This line of sight intersects a molecular cloud
associated with the Gum Nebula, an expanding supershell of radius
$\sim 70-130\,$pc \citep{yam99}.  The CO gas associated with
the \spitzer\ field has negative radial velocities, placing it on the
near side of the bubble.  Since the expansion center is at most
500\pc\ from the Sun \citep*{woe01}, we estimate a distance $d\sim
400\pc$ to our field.  The 24\micron\ and 70\micron\ fields also include
the object DC 254.5--9.6 from the optical catalog of
southern dark clouds of \citet{har86}. 

Before estimating the power spectrum, the \spitzer\ images shown in the
left column of Figure \ref{fig1} were processed to remove 
point sources and instrumental noise.  The emission from point sources
was characterized and removed using the StarFinder code written in
IDL \citep{di00a,di00b}.  
The images were then filtered to remove instrumental noise using a
multiresolution wavelet technique \citep{sta98}.  
The method required an estimate of the noise spatial response.
Following \citet*{mi03b}, we used the uncertainty images automatically
produced by the post-BCD pipeline to account for spatial variations in
the noise.  The right column of Figure \ref{fig1} shows the left
column images after point source removal and noise filtering.  

Power spectra for each of the images in Figure \ref{fig1} were
computed by squaring the amplitudes of the image fourier transforms.  
Prior to transforming the images, the outer 5\% of pixels were
multiplied by a cosine taper.  This apodization function minimized edge
effects caused by our non-periodic data.  In fourier wavenumber ($k$)
space, the effect is a slight smoothing of the amplitudes.  The radial
power spectra were derived by azimuthally averaging the squared
fourier amplitudes in equally-spaced wavenumber bins. Statistical
uncertainties were calculated as $1/\sqrt{n_s}$ times the azimuthal
average, where $n_s$ is the number of amplitude samples comprising the
average.  To ensure that we were not adding structure by noise filtering, 
the insignificant wavelet coefficients and scales that had been
rejected in the filtering process were summed to make a ``noise image.''  Power
spectra of the noise images showed no systematic trends as a function of
spatial wavenumber, consistent with uncorrelated ``white'' noise.  In
addition, a subregion of the \spitzer\ mosaics with no detectable
emission showed the same power spectrum as the noise images.   

\subsection{\iras\ Sky Survey Atlas Images}
We used surface brightness maps at 12,
25, 60, and 100\micron\ from the \iras\ Sky Survey Atlas (ISSA).  ISSA
maps were produced that covered
a $12\arcdeg\times 12\arcdeg$ region centered on the \spitzer\ subregions.
These maps were corrected for gain and offset variations by comparing
with measurements taken with the Cosmic Background Explorer (COBE)
Diffuse Infrared Background Experiment (DIRBE) \citep*{miv04}.  Before
power spectral estimation, the \iras\ data were processed for source
removal, noise filtering, and PSF deconvolution according to the
method described by \citet*{miv02}.   

\section{Results}
\subsection{\spitzer\ and \iras\ colors}
Table \ref{colortable} lists mean surface brightness ratios
$I_{\lambda}/I_{100}$ between the \spitzer\ and \iras\ data.  The point
source-subtracted \spitzer\ maps were convolved with the \iras\ beam and
resampled onto the ISSA grid.  Linear fits of $I_{\lambda}$ as a
function of $I_{100}$ were performed. The slopes and their errors are
reported in Table \ref{colortable}.   For most wavelengths the data
were well correlated with $I_{100}$.  After processing, the MIPS
70\micron\ image (Figure \ref{fig1} [f]) still showed systematic
instrumental features, most notably vertical stripes in the direction
of the scan legs.  We do not analyze the 70/100 color here, in
expectation of improved processing of the 70\micron\ data.  The
\spitzer -derived spectral energy distribution of the cirrus will be
discussed more fully in a future paper.  

\subsection{Spatial Power Spectrum}

\begin{figure}
\plotone{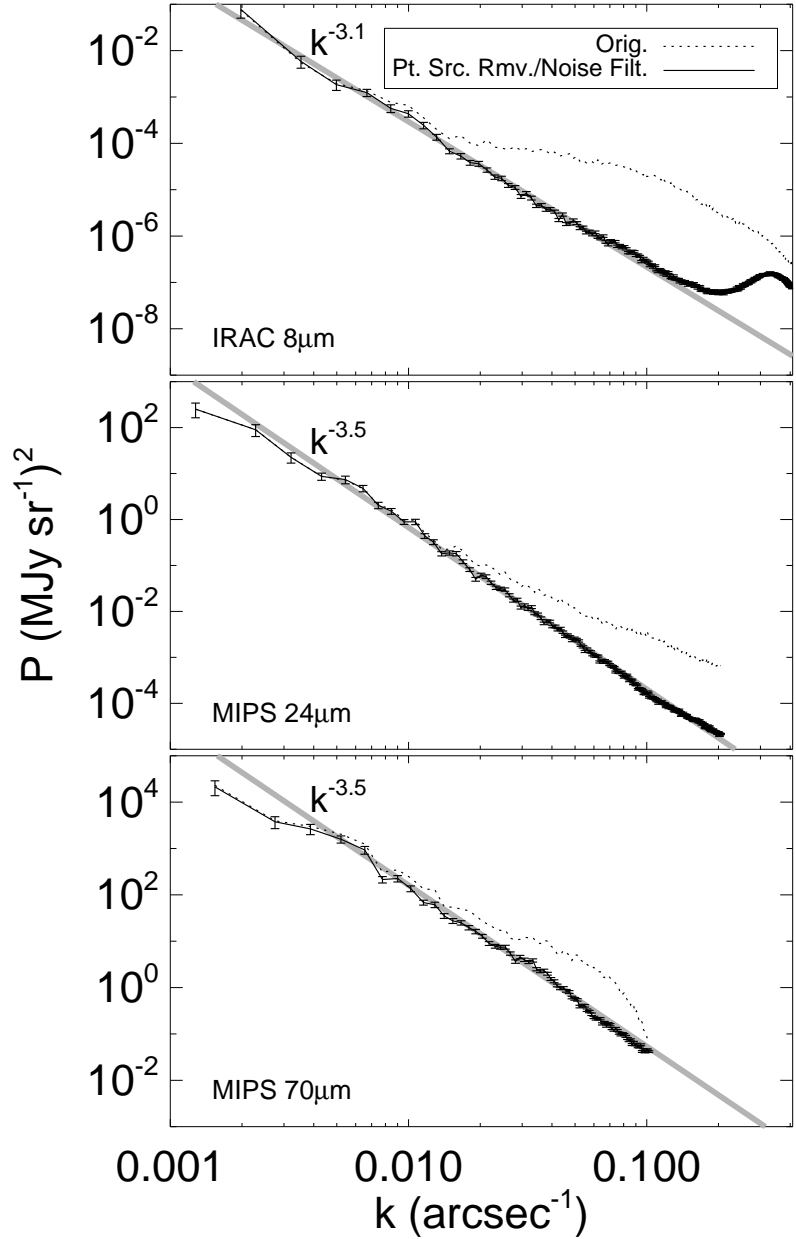}
\caption{Spatial power spectra of emission towards the {\sl Spitzer}
fields in Figure \ref{fig1}:  ({\it top}) IRAC 8$\micron$; ({\it middle})  MIPS 24$\micron$;
({\it bottom}) MIPS 70$\micron$.  Spectra correspond to the two stages of image
processing depicted in Figure \ref{fig1}:  original post-BCD mosaic
(dotted lines), and point source removed/noise-filtered (solid lines).
Power law fits to the point source removed/noise-filtered data are
superimposed as gray lines and labeled on the graphs. Statistical
error bars are plotted for the point source removed/noise-filtered measurements.
\label{fig2}}
\end{figure}

Figure \ref{fig2} displays the spatial power spectra of emission in
the fields shown in Figure \ref{fig1}.  In each of the three \spitzer\
data sets (IRAC 8\micron, MIPS 24 and 
70\micron) point source subtraction and noise filtering had an obvious
effect on the power spectra.  Both processing steps removed a
mostly flat ``white'' noise component.  The benefits of point source
removal were most pronounced in the IRAC 8\micron\ data, while noise
filtering improved greatly the linearity of the spectra (in log-log
plots) in all three fields.  

Power law fits ($P = Ak^{-\beta}$) to the point source removed/noise
filtered spectra are superimposed on the data in Figure \ref{fig2}.
The 24\micron\ and 70\micron\ fits gave the same result within the
formal fit errors, $\beta_{24} = -(3.52\pm 0.01)$ and $\beta_{70} =
-(3.48\pm 0.04)$.  In log-log plots, the 24\micron\ spectrum did not
deviate from a straight line for almost two decades of spatial frequency,
down to the spatial resolution of $\sim 5\arcsec$.  At the $\sim 400\pc$
distance of the emitting material, this is equivalent to a size scale of
$\sim 0.01\pc$.  

In contrast, the 8\micron\ image had a power
spectrum over the same spatial scales that was flatter by 0.4,
$\beta_{8} = -(3.13\pm 0.03)$, indicating more power at small scales
than the other two images.  A bump in the 8 \micron\ power
spectrum at high wavenumber was probably caused by incomplete
source subtraction.  Some sources were clearly not pointlike and
subtracting a scaled point spread function added high frequency
structure.  Unfortunately, this occured where the IRAC 8\micron\ image
might have provided additional information on the structure at the
smallest scales.  

\begin{figure}
\plotone{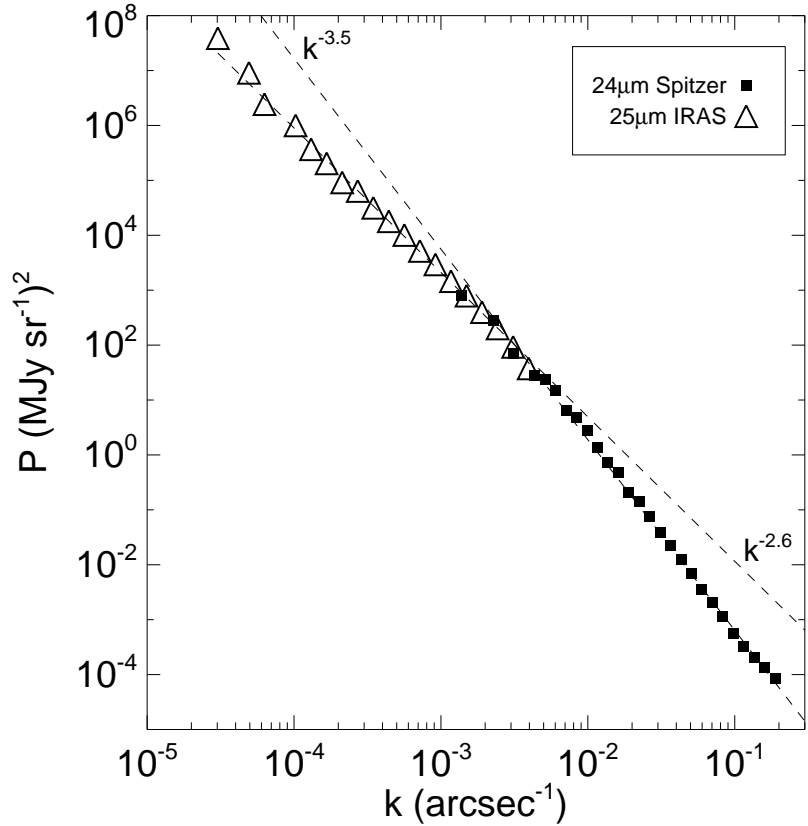}
\caption{Spatial power spectra at 25\micron\ ({\it open triangles}; {\sl IRAS}) and 
24\micron\ ({\it filled squares}; \spitzer) for the field encompassing the images
displayed in Figure \ref{fig1}.  Both spectra have been point
source-removed and noise-filtered, albeit with different methods (see
\S2).  The 25\micron\ spectrum has been normalized to intersect the
24\micron\ spectrum.  
The data have been rebinned in equally-spaced
logarithmic intervals for clarity.  Separate power law fits are
superimposed as dashed lines on the two spectra, and labeled on the
graph.  We do not plot statistical error bars since they would be
smaller than the symbols.
\label{fig3}}
\end{figure}

We extended the power spectrum of our 24\micron\ field to large scales
using the power spectrum of the \iras\ map that includes it.  Figure
\ref{fig3} shows a comparison of the MIPS 24\micron\ power spectrum
with that of an ISSA 25\micron\ image that includes the same field.
The 25\micron\ data were normalized (multiplied by a constant) to
match the 24\micron\ spectrum in the overlap region and all data
points were rebinned for clarity (except for the low-$k$ data,
each plot symbol is an average over more than one radial point).   The
25\micron\ power spectrum was fit by a power law with $\beta_{25} =
-(2.63\pm 0.02)$, which is shallower than the 24\micron\ spectrum by $\sim 1$.  Our
data suggest that there is a transition in the power law exponent from
--2.6 to --3.5 somewhere in the wavenumber range $2\times\expo{-3} < k
< 4\times\expo{-3} \,({\rm arcsec})^{-1}$.  This also seems to occur
in the 70\micron\ data.  If the decrease in slope is
real, then it is comparable to the bend in the power spectrum of 
\hi\ seen by \citet*{elm01} in the Large Magellanic Cloud (LMC), where the
spectral exponent also decreased by $\sim 1$ from low to high $k$.
This is predicted to occur when the transverse size of an image is
greater than the line of sight depth of the emitting medium, $d_{\litl
  LOS}$.  More precisely, the transition occurs at wavenumber $k =
1/2d_{\litl LOS}$ \citep[for theoretical and numerical support for this argument,
  see][]{laz00,mi03b}.  Under this hypothesis, we estimated
$d_{\litl LOS}$ for the 25 and 24\micron -emitting dust cloud.  Assume
that the transition occurs at $k_{\litl trans} = 0.003
\,({\rm arcsec})^{-1}$, or 1.54$\pc^{-1}$ if the distance is 400\pc. Thus, 
$d_{\litl LOS} = 1/2k_{\litl trans} = 0.3\pc$.  Without
invoking distance we can use the angular size of our map,
$\Theta_{\rm map}$, and the transition wavenumber to estimate the
size-to-thickness ratio of the medium, $L/d_{\litl LOS} =
2\Theta_{\rm map} k_{\rm trans}.$  Taking $\Theta_{\rm map} = 0 \pdeg
3$, \ie\ the size of the 24\micron\ image used to compute the
high-$k$ power spectrum, $L/d_{\litl LOS} = 6.5$.  This is
probably a lower limit, since the emission continues beyond the edges
of the map.

\section{Discussion}
The \spitzer\ 24\micron\ and 70\micron\ surface brightness maps of 
the Gum Nebula have the same power
spectral index, $\beta \approx -3.5$.  The 24\micron\ data in particular
conform well to a --3.5 power law down to the spatial
resolution of $\sim 5\arcsec$, equivalent to a size scale of $\sim
0.01\pc$ at the estimated distance of
$\sim 400\pc$.  Thus, {\it the self-similar structure of the
  24\micron\ emission continues to milliparsec scales.}  
To understand properly the relationship between the emission and the column
density, we need to model the variations in grain heating and
emissivity as a function of density, extinction, and distance from ultraviolet sources.
Nevertheless, these $\beta$ estimates that
we have derived for the mid-infrared dust surface brightness are close to
the value $-3.6$ reported for the density and velocity fields of 
\hi\ in emission in the Galaxy and the LMC
\citep[see][respectively]{mi03a,elm01}.  A spatial power spectrum
index of $-3.5$ is also close to the prediction $\beta = -11/3$, based
on the incompressible turbulence theory of \citet{kol41}.  In
contrast, the velocity integrated emission from dense molecular gas 
typically yields values of $\beta \sim -2.8$ \citep{stu98,ben01,pad04}.

Our power spectral analysis may have enabled us to access
the depth of 24\micron\ emitting material along the line of sight, $d_{\litl
  LOS} \approx 0.3\pc$.  
The steepening of the power spectrum from $\beta_{\rm 2D}$ to 
$\beta_{\rm 3D} \approx \beta_{\rm 2D}-1$ 
(in our case from --2.6 to --3.5) is predicted to
occur when the size of the map is larger than $d_{\litl LOS}$
\citep{mi03b}.  
At small wavenumbers the image power spectrum reverts to that of
a 2-D field, \ie\ the projected 3-D volume statistics are equivalent
to those of a 2-D slice, and the power spectrum flattens.\footnote{An
alternative interpretation holds that the break in the spectral index
is caused by the lack of self-similarity in the 3-D turbulence above the scale where the
system becomes two-dimensional \citep*{pad01}.  Even under this
interpretation, however, the observed break would mark the thickness
of the system.}  
The 2D/3D transition has been observed before in the
LMC \citep{elm01}, but this is the first example in the Milky Way.

What is the physical significance of the 0.3\pc\ emitting layer
thickness?  There are two possibilities that depend on the excitation
of dust grains into emission in the
24\micron\ band.  If the emission is dominated by dark regions, then
$d_{\litl LOS}$ could be the average ``skin
depth'' for penetration of UV photons, or the depth beyond which the
abundance of emitters decreases drastically.  The skin depth effect
should be even more apparent in the 8\micron\ emission, but less apparent
in the 70\micron\ emission.

If, on the other hand, the emission comes from mostly diffuse regions
with $\av\lesssim 1\,$mag, then UV photons permeate the entire cloud
and $d_{\litl LOS}$ is the average thickness of the Gum Nebula shell
itself.  While opaque lines of sight do exist in the Gum Nebula
\citep[\eg\ the cometary globules studied by][]{sri92}, on $\sim
2\arcmin$ scales our field has only moderate extinction, $A_{\rm v} <
1.1\,$mag \citep[derived from the brightest $^{12}$CO contour
of][]{yam99}.  Furthermore, the 70\micron\ power spectrum also shows
evidence of a break at $k = 0.003\,({\rm arcsec})^{-1}$, implying that
$d_{\litl LOS} = 0.3\pc$ is independent of excitation.  The 24 and
70um data show only one scale where the 2D-3D transition occurs.


It would not be surprising if the 24\micron\ size-to-thickness ratio,
$L/d_{\litl LOS} \gtrsim 6.5$, applied not only to the small portion
of the Gum Nebula that we have observed, but to the diffuse medium of
the Galaxy in general.  In an \hi\ survey of the cold neutral medium,
\citet{hei03} demonstrated that clouds are ``sheetlike,'' with
size-to-thickness ratios of up to 280.  Far from being unusual, the
2D/3D power spectrum break that we see may be a normal feature of
large spatial dynamic range observations of the local cold neutral ISM.

\acknowledgments

    This work is based on observations made with the Spitzer Space
    Telescope, which is operated by the Jet Propulsion Laboratory,
    California Institute of Technology under NASA contract
    1407. Support for this work was provided by NASA through an award
    issued by JPL/Caltech.  We thank the referee, Paolo Padoan, for
    his comments that improved the manuscript.

\clearpage



\clearpage


\clearpage
\begin{table}
\tablewidth{300pt}
\begin{center}
\caption{Colors in {\sl Spitzer} and {\sl IRAS} Maps \label{colortable}}
\begin{tabular}{rc}
\tableline\tableline
\multicolumn{1}{c}{$\lambda$} & \multicolumn{1}{c}{$I_{\lambda}/I_{100}$\tablenotemark{a}} \\
\multicolumn{1}{c}{($\micron$)} &  \\
\tableline

8 & $0.039\pm 0.005$  \\
12 & $0.029\pm 0.005$ \\
24 & $0.041\pm 0.005$ \\
25 & $0.048\pm 0.003$ \\
60 & $0.315\pm 0.003$ \\
\tableline
\end{tabular}
\tablenotetext{a}{Slope of linear fit to the surface brightness
  $I_{\lambda}$ as a function of $I_{100}$, $\pm$ formal fit errors.}
\end{center}
\end{table}

\end{document}